\documentclass[aps,prb,twocolumn,longbibliography]{revtex4-1}

\usepackage{hyperref,xcolor,ragged2e,array}

\newcommand{\QUOTE}[2]{\begin{itemize}\item[]\textit{``#1"}\end{itemize}}
\newcommand{\Quote}[1]{\begin{itemize}\item[]#1\end{itemize}}

\newcommand{\ssout}[2]{#2} 
\newcommand{\SSOUT}[2]{} 

\newcommand{\demtable}{
\begin{table}[t]\caption{\label{tab:demographics}Demographic breakdown of students who completed Foundations from 2014 to 2016 (about 22 per year), and students who earned a physics bachelor's degree from CU between 2012 and 2016 (about 38 per year). Data were provided by the CU Office of Institutional Research.}
\begin{ruledtabular}
\begin{tabular}{lrr}
 & \multicolumn{1}{r}{Course (\%)} & \multicolumn{1}{r}{Dept. (\%)}\\ \hline
Men & 76 & 84   \\
Women & 24 & 16   \\
White & 52 & 79  \\
Asian American & 14 & 6  \\
Latinx or Hispanic & 11 & 4 \\
Black or African American & 3 & $<1$  \\
Native American or Alaska Native & 6 & 0 \\
Native Hawaiian or Pacific Islander & 0 & $<1$ \\
Other or unknown race or ethnicity & 0 & 5  \\
International students & 15 & 5 \\
First-generation college student & 15 & 12 
\end{tabular}
\end{ruledtabular}
\end{table}
}

\newcommand{\codetable}{
\begin{table}[t]\caption{\label{tab:codes}GRF response coding results for Emily ($N=63$), Taylor ($N=71$), and both instructors ($N=134$).}
\begin{ruledtabular}
\begin{tabular}{lrrr}
Code category & Emily (\%) & Taylor (\%) & Both (\%) \\ \hline
Encouraging statements & 97 & 72 & 84 \\
Normalizing statements & 73 & 38 & 54 \\
Empathizing statements & 49 & 11 & 29 \\
Strategy suggestions & 44 & 72 & 59 \\
Resource suggestions & 19 & 24 & 22 \\
Feedback on structure & 44 & 45 & 45
\end{tabular}
\end{ruledtabular}
\end{table}
}

\begin{document}


\title{Personalized instructor responses to guided student reflections: \\ Analysis of two instructors' perspectives and practices}

\author{Daniel L. Reinholz}
\email{daniel.reinholz@sdsu.edu}
\affiliation{Department of Mathematics \& Statistics, San Diego State University, San Diego, CA 92182, USA}

\author{Dimitri R. Dounas-Frazer}
\affiliation{Department of Physics, University of Colorado Boulder, Boulder, CO 80309, USA}

\date{\today}

\begin{abstract}
One way to foster a supportive culture in physics departments is for instructors to provide students with personal attention regarding their academic difficulties. To this end, we have developed the Guided Reflection Form (GRF), an online tool that facilitates student reflections and personalized instructor responses. In the present work, we report on the experiences and practices of two instructors who used the GRF in an introductory physics lab course. Our analysis draws on two sources of data: (i) post-semester interviews with both instructors and (ii) the instructors' written responses to 134 student reflections. Interviews focused on the instructors' perceptions about the goals and framing of the GRF activity, \ssout{}{and} characteristics of good or bad feedback\ssout{, and impacts of the GRF on the nature of teacher-student relationships}{}. Their GRF responses were analyzed for the presence of up to six types of statement: encouraging statements, normalizing statements, empathizing statements, strategy suggestions, resource suggestions, and feedback to the student on the structure of their reflection. We find that both instructors used all six response types, \ssout{and that they both perceived that the GRF played an important role in the formation of meaningful connections with their students}{in alignment with their perceptions of what counts as good feedback}. This exploratory qualitative investigation \ssout{}{demonstrates that the GRF can serve as a mechanism for instructors to pay personal attention to their students. In addition, it} opens the door to future work about the impact of the GRF on student-teacher \ssout{relationships}{interactions}.
\end{abstract}

\maketitle


\section{Introduction}

Reflection is an important skill in learning physics,\cite{Ward2014,Mason2010,Etkina2010,Scott2007,May2002} and is a key part of learning more generally.\cite{NASEM2015,Zimmerman2002} Previously we have described how structured reflection activities can augment physics courses that focus on iterative improvement of models\cite{Reinholz2016} and apparatuses.\cite{Gandhi2016} We have also developed an online tool, the {Guided Reflection Form (GRF)}, that facilitates student reflection and personalized instructor responses\ssout{.\cite{Dounas-Frazer2015}}{. The GRF was designed to support students in describing a past experience, setting a goal for improvement, and identifying specific steps for achieving that goal.} In a study of the GRF, we focused on the structure of students' reflections in a physics course for future teachers\ssout{; students in that study successfully used the GRF to narrate specific experiences upon which they wanted to improve and articulate goals and/or action plans for improvement}{}.\cite{Dounas-Frazer2015} In this article, we explore the GRF activity in a different context and from a different perspective. Here, we focus on how the GRF was implemented in an introductory lab course, and we characterize the types of feedback that the two instructors of that course provided in response to their students' reflections.

Our analysis of instructors' responses to students' reflections is motivated by an overarching desire to cultivate a culture of support and inclusiveness in undergraduate physics courses. In particular, we aim to develop and implement research-based educational tools that may counter weed-out culture. We consider weed-out culture to be a set of traditional educational practices and beliefs aimed at sorting and selecting the students seen as most capable, while ``weeding out" the rest (i.e. removing them from the system). In their landmark study of undergraduate student attrition from science, mathematics, and engineering majors, Seymour and Hewitt described the disproportionate impacts of this culture on marginalized groups:\cite{Seymour1997}
\Quote{
The most serious criticisms of the weed-out system, however, focused on its disproportionate impact on men of color and on all women. Even well-prepared, these two groups tend to enter basic classes feeling uncertain about whether they `belong.' The loss of regular contact with high school teachers who encouraged them to believe in their ability to do science exposes the frailty of their self-confidence. Faculty who teach weed-out classes discourage the kind of personal contact and support which was an important part of high school learning. It is, as some students describe, a `weaning away' process by which faculty transmit the message that it is time to grow up, cast aside dependence on personally-significant adults and take responsibility for their own learning. This attitude is perceived by students in the reluctance of teachers to answer questions, brusqueness in response to `trivial' inquiries, failure to offer praise or encouragement, disinclination to discuss academic difficulties in a personal manner, carelessness in keeping office hours, and a `no excuses' stance on test results. The difficulty of getting personal attention was troubling to many students, but it was especially troubling to those whose presence in [science, mathematics and engineering] classes was the result of considerable personal attention and encouragement by particular high school teachers. (p.~132)
}
The GRF was designed to provide avenues of communication through which instructors and students can engage in precisely those interactions that are discouraged by weed-out culture. As educators ourselves, the authors of the paper have used the GRF to this end in multiple contexts. In the current study, our goal was to understand the extent to which the GRF opens up such opportunities for other instructors, particularly those who were not involved in the iterative design process through which the GRF was developed. Indeed, as we will show, both instructors in our study \ssout{perceived the GRF as valuable for developing personally-significant relationships with their students, and both used it}{used the GRF} to provide their students with personal attention and encouragement. 

When imagining how instructors might ideally use the GRF to foster supportive student-teacher \ssout{relationships}{interactions}, Brown's metaphor of ``sitting on the same side of the table"\cite{Brown2012} is helpful. Brown drew on this metaphor to create a checklist for feedback that includes items like sitting ``next to you rather than across from you," putting ``the problem in front of us rather than between us (or sliding it toward you)," and modeling ``the vulnerability and openness that I expect to see from you"~(p.~204). After outlining her checklist, she asked,
\Quote{
How would education be different if students, teachers, and parents sat on the same side of the table? How would engagement change if leaders sat down next to folks and said, ``Thank you for your contributions. Here's how you're making a difference. This issue is getting in the way of your growth, and I think we can tackle it together. What ideas do you have about moving forward? What role do you think I'm playing in the problem? What can I do differently to support you?" (pp.~204--205)
}
The image of two people sitting on the same side of the table inspires our vision for how the GRF could shape classroom practices in physics: instructors and students working side-by-side to tackle academic problems together\ssout{, building meaningful student-teacher relationships along the way}{}.

We present a qualitative exploration of two instructors' implementations of the GRF in a lab course for first-semester undergraduate students interested in majoring in physics. The instructors were physics graduate students, and the course was designed and offered as part of a student-led diversity initiative in the instructors' physics department.  We conducted hour-long post-semester interviews with both instructors, and we collected electronic copies of 134 student reflections and corresponding instructor responses that were generated via the GRF. Using these data, we construct a rich picture of each instructor's unique implementation.

This paper is organized as follows. In Sec.~\ref{sec:background}, we describe the GRF activity and provide a brief overview of some of the literature on feedback. We describe the programmatic and course context for our study in Sec.~\ref{sec:context}, outline our research methods in Sec.~\ref{sec:methods}, and present results from our analyses of instructors' interviews and GRF responses in Sec.~\ref{sec:results}.  Finally, in Sec.~\ref{sec:discussion}, we summarize our findings, discuss their implications, and identify potential future directions for research and development of the GRF.


\section{Background}\label{sec:background}

We begin our discussion by describing the GRF and summarizing relevant literature about instructor feedback practices. When describing the GRF, we focus on how it has typically been used in other courses we have taught and/or studied. The instructors in the present study deviated slightly from this typical usage, as discussed in Sec.~\ref{sec:context}.


\subsection{Guided Reflection Form}\label{sec:GRF}
The GRF has been described in detail elsewhere,\cite{Dounas-Frazer2015} so we provide only a brief overview here. The GRF is an online tool, similar to a survey, that provides questions and other prompts to guide student reflections about issues of resilience, collaboration, and organization. Once per week, students are tasked with submitting a reflection via the GRF. Reflections may focus on any aspect of the students' learning experience, whether or not it is directly related to the course in which the GRF is being implemented. Instructors then read the reflections and provide individualized responses to each student based on the content of their (the students') reflection. This cycle of reflection and feedback repeats, ideally facilitating an ongoing written dialogue between each student and the instructor.

\ssout{}{Student responses can be collected by having students complete an online survey or submit individual electronic documents. In the former case, instructors can export student reflections into a spreadsheet, write their responses in the spreadsheet, and then use a mail merge program to generate individual documents with student responses and corresponding instructor feedback. In the latter case, the instructor can write their feedback directly on the submitted document.  Based on our own experiences using the GRF, responding to reflections takes about 3 to 5 min per student. Instructors in this study reported spending about 10 to 15 min per student responding to reflections.} In large classes, the time required for an instructor to respond to each student individually can be prohibitively large; hence, this activity is most suitable for classes with 10 to 20 students \ssout{}{per instructor or, in larger courses, per teaching or learning assistant}. The GRF has been implemented in a variety of contexts, including high school-level computer science and upper-division quantum mechanics.

When using the GRF, students are presented with a prompt instructing them to recall \ssout{a scenario}{an experience} from the previous week upon which they would like to improve. \ssout{}{Such experiences could include, for example, procrastinating on a long-term project.\cite{Dounas-Frazer2015}} Next, they are asked to choose one of three focus areas for reflection: bouncing back from failure or other setbacks; building a network and developing collaboration skills; or becoming an organized, self-aware, and mindful person. For students who would prefer to write about a different topic, the GRF includes a fourth option for ``something different." After students choose a topic for their reflection, the GRF presents a short paragraph describing the importance of the skills related to the topic. Regardless of topic, students are asked to write short responses to two reflection prompts:
\begin{itemize}
\item Describe the specific experience from last week that you would like to improve upon.
\item Describe an aspect of this experience that you can improve in the future. (Provide at least one concrete strategy that you will use to become more successful.)
\end{itemize}
The GRF prompts were designed with three aspects of reflection in mind: students should (i) revisit a salient experience from the previous week, (ii) set a future goal for improvement, and (iii) articulate specific steps for achieving that goal. In a study of undergraduate students using the GRF in a pedagogy course for future physics teachers, we found that all students successfully used the GRF to engage in multiple aspects reflection.\cite{Dounas-Frazer2015} In this article, we explore for the first time the ways that instructors use the GRF to provide feedback to their students.


\subsection{Feedback}\label{sec:feedback}

Providing feedback to students has a significant impact on their learning, but not all feedback is equally useful. For instance, there are a number of characteristics that make feedback effective, including specificity and timeliness.\cite{shute2008} Process-level feedback is particularly effective for enhancing learning; such feedback focuses on students' ability to strategize about their learning and to seek help when needed.\cite{Hattie2007} When students receive feedback about their learning strategies, it draws attention to the ways in which they can adapt to become more effective learners.\cite{Rattan2012} In contrast, praise can have unpredictable impacts---and can even inhibit learning, especially if it is perceived as undeserved---because it draws students' attention to themselves rather than the task at hand.\cite{Hattie2007}

A popular way of interpreting these findings is through the concept of mindset;\cite{Dweck2006} indeed, this concept informed the perspectives of one of the instructors in our study. Here, ``mindset" refers to students' beliefs about the nature of intelligence. Mindset is commonly described using a fixed/growth dichotomy: in the extreme cases, people with a fixed mindset view intelligence as static and unchangeable beyond a predetermined level, whereas those with a growth mindset view intelligence as malleable and something that can be improved with effort.\cite{Dweck2006} Using the language of mindset, providing process-level feedback is consistent with a growth mindset.\cite{Rattan2012} In particular, feedback that emphasizes self-improvement can bolster students' beliefs in their own capability to succeed.\cite{Bandura1997} On the other hand, praising a student for being ``smart" may reinforce a fixed mindset,\cite{Dweck2006} and feedback that communicates a lack of faith in a student's capabilities can undermine their confidence, motivation, and willingness to attempt challenging tasks.\cite{Bandura1997}

\ssout{}{
From this literature, we infer two principles that could support instructors' effective use of the GRF:
\begin{enumerate}
\item[P1.] Praise should should focus on students' efforts to improve, express confidence in their ability to improve, and be sincere.
\item[P2.] Process-level feedback should identify specific areas for improvement and suggest strategies that students can use to improve their learning.
\end{enumerate}
Each principle is also informed by a particular aspect of weed-out culture, as described by Seymour and Hewitt.~\cite{Seymour1997} In particular, they identified instructors' ``failure to offer praise" and ``disinclination to discuss academic difficulties in a personal manner" as factors contributing to weed-out culture. However, because not all forms of praise support student perseverance, P1 recommends a particular process for giving praise. Similarly, P2 can be thought of as a guideline for how instructors can discuss academic difficulties with students. Taken together, these two principles align with part of Brown's vision for students and teachers sitting on the same side of the table.~\cite{Brown2012} For example, should a teacher say to a student, ``This issue is getting in the way of your growth, and I think we can tackle it together," they would be identifying an area for improvement (P2) and expressing confidence in the student's ability to improve (P1).
}

Importantly, feedback must be understood in the context of the learning environment in which it is given and received. For instance, feedback is more effective when instructors create classroom communities that normalize failure and value criticism. Students in these settings are better situated to receive and use feedback.\cite{Brookhart2008} Therefore, one way for instructors and students to sit on the ``same side of the table" is to be embedded in a culture where students are in the habit of receiving timely, sincere, and critical feedback focused on their strategies for self-improvement. In the sections that follow, we describe a course in which the instructors aspired to foster such a culture, in part by using the GRF.


\section{Context}\label{sec:context}

\ssout{}{Our study is an exploratory qualitative investigation of two instructors' feedback practices. As Eisenhart argues,\cite{Eisenhart2009} such qualitative studies are responsible for ``providing sufficient detail about the researched context for a person with intimate knowledge of a second context to judge the likelihood of transferability."~(p. 56). Accordingly, we describe the context for our study at three grain sizes: organization, course, and activity.}

\subsection{Organizational context}

The two instructors in our study---Emily and Taylor---were both physics graduate students at the University of Colorado Boulder (CU), a predominantly white public R1 university with a large physics program. Emily was a white woman and Taylor was a white man. They co-taught a course called Foundations of Scientific Investigation (hereafter ``Foundations"). Foundations was designed as part of a student-led organization called CU-Prime. CU-Prime is a member of The Access Network (hereafter ``Access").\footnote{The Access Network, \url{http://accessnetwork.org}} Access organizations---including The Berkeley Compass Project,\cite{Albanna2013} The Chi-Sci Scholars Program,\cite{Sabella_inreview} and several other organizations---are characterized by student leadership and a commitment to improving diversity in the physical sciences through community building. To achieve their goals, these organizations offer multiple services designed to support students from underrepresented groups and raise awareness about issues of marginalization in physics. Examples of services include summer programs,\cite{Dounas-Frazer2013TPT} diversity workshops,\cite{Dounas-Frazer_inpress} mentorship programs,\cite{Zaniewski2016} and courses with multi-week final projects.\cite{Gandhi2016} \ssout{}{M}any of the courses designed and run by Access organizations use the GRF or similar tools to facilitate cycles of student reflection and instructor feedback. \ssout{}{In this work, we focus on the implementation of the GRF in Foundations.}

\subsection{Course description}

Foundations was first designed and taught in 2014 and was subsequently refined and taught in 2015 and 2016. It is a 14-week, fall-semester course designed for first-year undergraduate students interested in majoring in physics. On average, 22 students enroll in Foundations each semester. Students from underrepresented and/or minority racial and gender groups are especially encouraged to participate in the course; a demographic breakdown of students who completed the course is provided in Table~\ref{tab:demographics}. The overarching goals of Foundations are twofold: build community among students enrolled in the course, and introduce students to the practice of research. Two corresponding subgoals are for students to practice developing theoretical models of scientific phenomena, and to reflect on and refine their coursework in Foundations and other courses.

Each semester, the Foundations class met twice weekly for 75 minutes per meeting, and the course consisted of 2 successive 7-week halves. Consistent with the subgoals of the course, each half included both experimental activities that focused on building models as well as activities that engaged students in the practice of reflection.  During the first half of the course, students worked in groups on a set of guided optics experiments. They also used the GRF to reflect on their collaboration, organization, and resilience. During the second half, students worked in groups on multi-week final projects under the guidance of graduate student mentors; a similar approach to final projects has been described elsewhere.\cite{Gandhi2016} In this part of the course, students used a tool similar to the GRF to reflect on goals, challenges, and successes related to their final projects.

\demtable

Since its inception, Foundations has been divided into 2 parallel sections of about 10 students. Each section has been co-taught by 2 instructors, for a total of 4 instructors per semester. By design, each co-teaching pair has been mixed gender and has comprised one undergraduate student and one graduate student. Most instructors have only taught the course once, and former instructors meet with new instructors during the summer to discuss teaching strategies for the upcoming fall semester.  Emily and Taylor taught Foundations concurrently, but in different sections. Hence, each was a member of a co-teaching pair, but neither was the other's co-teacher.

\subsection{GRF activity}

We introduced Emily, Taylor, and their co-teachers to the GRF during the summer before they started teaching the course. Based on discussions between the authors and the instructors, the instructors' implementation of the GRF differed from that described in Sec.~\ref{sec:GRF} in two ways. First, the GRF was assigned only during the first half of the course. This choice was made because a different reflection tool was deemed more appropriate for the second half of the course. Second, GRF focus areas were assigned by the instructors. Reflections focused on collaboration during weeks 1 and 2, organization during weeks 3, 4, and 5, and resilience in weeks 6 and 7. This choice was informed in part by the anticipated progression of \ssout{}{student-to-student} relationships in the course. Students would still be getting to know their group members during the first couple weeks of the semester, and they might not feel comfortable sharing about their experiences of failure with their instructors until several weeks had passed. This choice was also informed by the fact that many introductory courses have multiple midterms, making it important to develop good time management practices as early as possible. {Although students were not graded on the quality of their reflections, they were awarded a small amount of course credit for completing the GRF activity.}

At the start of the fall semester, we provided Emily and Taylor with guidelines for giving feedback.\cite{GRFguidelines} The guidelines were based on our experiences with the GRF\cite{Dounas-Frazer2015} and a precursor to the GRF,\cite{Gandhi2016} and they emphasized the importance of communicating the goal of the activity to students as well as providing feedback on both the structure and content of students' reflections. Based on the instructors' internal decisions about division of labor, Emily and Taylor were each solely responsible for responding to all the reflections written by students in their respective sections; their co-teachers did not provide any individualized written feedback to students via the GRF activity. In this paper, we explore the ways that Emily and Taylor incorporated the GRF into their teaching. \ssout{}{In doing so, we aim to provide clear examples of instructor feedback, as facilitated by the GRF. These examples could inform future instructors' use of the GRF in other  contexts.}


\section{Methods}\label{sec:methods}

\ssout{This}{There are multiple ways in which the goals of the GRF, the Foundations course, and the CU-Prime organization are theoretically aligned. One line of reasoning is as follows: Seymour and Hewitt noted that weed-out culture has a ``disproportionate impact on men of color and all women;"\cite{Seymour1997} these populations are better represented in Foundations than in the CU Physics Department as a whole (Table~\ref{tab:demographics}); the GRF was designed to counter some aspects of weed-out culture; and, finally, the Foundations course was designed to support CU-Prime's commitment to improving diversity in physics. Thus, implementing the GRF in Foundations is in alignment with the diversity-oriented mission of CU-Prime. More narrowly, the GRF directly aligns with the course goal of engaging students in the practice of reflection. However, our present focus is not on student experiences or outcomes as they relate to improving diversity in physics. Rather, we are interested in teacher practices. Accordingly, this} study is a qualitative exploration of the ways that Emily and Taylor implemented the GRF in the Foundations course.

We conducted post-semester interviews with both Emily and Taylor, focusing on their \ssout{goals for, perspectives on, and}{} engagement with the GRF activity. To corroborate the instructors' self-reported response practices, we also collected and analyzed all instructor responses that were generated via the GRF. Thus, our study enables us to describe how Emily and Taylor implemented the GRF using their own words and authentic examples of the responses they provided to students. Our goal is not to make generalizable statements about either the GRF or instructors who use it, but rather to provide insight into the various ways that instructors might take up the GRF for use in their classrooms. In this section, we describe our data sources and analysis methods.


\subsection{Post-semester interviews}

At the end of the semester, we conducted semi-structured interviews with Emily and Taylor to gain insight into their perspectives on the GRF and other aspects of the course. Emily and Taylor were interviewed separately, each for about an hour.  Interviews focused in part on \ssout{three}{two} themes: the instructors' perceptions about the (i) goals and framing of the GRF activity, \ssout{}{and} (ii) characteristics of good or bad feedback\ssout{, and (iii) impacts of the GRF on the nature of teacher-student relationships}{}. We chose \ssout{the first two}{these} themes because they give us insight into why and how the instructors were using the GRF. \ssout{The third theme was chosen because it reflects a goal of the Foundations course that runs counter to weed-out culture: to foster supportive relationships among students and teachers.}{}

The first author transcribed both interviews, and the transcripts are the data that we analyzed. We collaboratively identified all excerpts that addressed the three themes that comprised the foci of our interviews. For each theme, we selected several representative excerpts and constructed two vignettes about the implementation of the GRF, one each for Emily and Taylor. These vignettes are presented in Sec.~\ref{sec:results}.


\subsection{GRF responses}

In total, 22 students were enrolled in Foundations: 10 in Emily's section and 12 in Taylor's. Each student was required to complete 7 reflections using the GRF. Of 154 possible GRF-based reflections, 135 were submitted. This corresponds to a completion rate of 88\%, which is consistent with the GRF completion rate observed in another study.\cite{Dounas-Frazer2015} The majority of students completed all or most reflections: 12 students completed all 7 reflections, 8 completed 5 or 6, and 2 completed 3 or 4. This distribution was roughly the same in both sections, resulting in similar completion rates for Emily's section (90\%) and Taylor's section (86\%). Each instructor responded only to reflections completed by students enrolled in their section. Almost every submitted reflection received a personalized response from either Emily or Taylor; only 1 reflection received no instructor response.

We analyzed both instructors' GRF responses using an \emph{a priori} coding scheme. This scheme was not directly informed by existing frameworks for effective feedback, such as those described in Sec.~\ref{sec:feedback}. Rather, our goal was to explore these data through an analytic lens informed by the language associated with the tool itself. Hence our scheme was directly informed by the guidelines\cite{GRFguidelines} we gave to Emily and Taylor at the start of the semester as well as a preliminary analysis of the instructors' feedback styles as self-reported during interviews. Based on our guidelines, we created code categories for normalizing statements, empathizing statements, resource suggestions, and feedback on the structure of the reflection. Based on our preliminary analysis of Emily's and Taylor's interviews, we created additional code categories for encouraging statements and strategy suggestions, respectively. Thus, our coding scheme included categories corresponding to 6 distinct types of response:
\begin{enumerate}
\item \emph{Encouraging statements} serve to motivate the student or validate their experiences and efforts. Examples include: ``You're doing great," ``I believe in you," ``You can do this," and, ``I'm glad you're using this strategy."
\item \emph{Normalizing statements} involve communicating to the student that what they are experiencing is normal, common, and/or unsurprising; this can be accomplished by relaying a personal anecdote or making an appeal to the general student experience. Examples include: ``I experienced something similar," and ``Lots of students go through this."
\item \emph{Empathizing statements} involve empathizing with the student cognitively, in a parallel emotional capacity, or in a reactive emotional capacity. Examples include: ``I understand where you're coming from," ``Your story makes me feel upset, too," and ``I'm excited that you are enjoying class."
\item \emph{Strategy suggestions} include both direct and indirect suggestions, the latter of which may take the form of anecdotes or questions. Examples include: ``You should use a day planner," ``When I was in this situation, I used a day planner," and ``Have you thought about using a day planner?"
\item \emph{Resource suggestions} also include both direct and indirect suggestions. Examples include: ``You should go to office hours," ``When I was in this situation, office hours were very helpful," and ``Have you thought about going to office hours?"
\item \emph{Feedback on reflection structure} focuses on the way the student wrote their reflection---such as whether the reflection provided enough detail or articulated a goal/strategy for improvement---and may be formulated as a comment or question. Examples include: ``Please write a longer reflection next week," and ``How can you achieve this goal?"
\end{enumerate}
\ssout{Some individual statements received two codes. For example, empathizing with a student by normalizing their feelings was a common strategy for Emily, and about half of her empathizing statements were also coded as normalizing statements.}{In terms of the principles for feedback outlined in Sec.~\ref{sec:feedback}, categories 1 to 3 map onto principle P1, which suggests that praise focus on students' efforts and be sincere. While student data would be needed to determine the perceived sincerity of instructor feedback, normalizing and empathizing statements could contribute to such perceptions. Categories 4 to 6 map onto principle P2, which recommends that process-level feedback suggest strategies for improvement.} 

We coded all 134 GRF responses via the following process. The second author read through each instructor response and identified all statements that aligned with one or more categories in our coding scheme. \ssout{}{Some individual statements received two codes. For example, empathizing with a student by normalizing their feelings was a common strategy for Emily, and about half of her empathizing statements were also coded as normalizing statements.} Then, for each category, the first author read through all the coded statements to verify that they matched the category definition, making note of any statements that did not fit the category. In total, such discrepancies were identified in only 10 responses; each of these discrepancies was reconciled through discussion among both authors. While we did not analyze student reflections, we read each reflection in order to provide context for the corresponding instructor response.

An initial version of our coding scheme included categories for additional types of statements, including praise for a student's intellect, achievement, or effort, as well as instances where an instructor articulated their expectations for student behavior. However, for each of these additional categories, we found few or no corresponding statements among the GRF responses in our dataset. Hence, these categories were discarded from our analysis.

Meanwhile, each of the 6 response categories in our final coding scheme appeared in at least 22\% of the 134 distinct responses (see Table~\ref{tab:codes}). Moreover, each of the responses included at least 1 statement corresponding to our code categories, and most responses comprised multiple types of statement. Indeed, 62\% of responses received at least 3 codes. This suggests that there was a good mapping between our \emph{a priori} coding scheme and our dataset. Nevertheless, the scheme was not comprehensive. For example, it did not capture instances where instructors used the GRF to communicate with students about certain aspects of Foundations (e.g., clarifying when homework is due or responding to schedule conflicts between Foundations and campuswide events).

One limitation of this analysis is that, due to the small number of students in each section, it is not possible to make strong claims about an instructor's feedback style. Consider, for example, a scenario where one section has many students who engage in the GRF in a meaningful way on their own, but the other section has many students who instead engage in a only cursory way. In this scenario, the responses written by the instructor of the former section may include relatively few instances of structure feedback compared to those of the other. Hence, differences in the frequency of particular types of feedback may be due to differences in student populations, not differences in the two instructors' response styles. Therefore, when discussing results in Sec.~\ref{sec:discussion}, we use  instructors' self-reported practices (i.e., interview data) to help interpret the results of our coding scheme. {In addition, Emily and Taylor read a draft of this manuscript, and both instructors indicated that they felt their perspectives and practices were accurately portrayed.}

In the following section, we report the results of our analyses of the interview data and the instructors' responses to the GRF.


\section{Results and interpretation}\label{sec:results}

\codetable

A summary of our GRF response coding is provided in Table~\ref{tab:codes}. For both instructors, encouraging statements were present in most response whereas resource suggestions were relatively sparse. In comparison to Taylor's responses, Emily's GRF responses were characterized by higher rates of encouraging, normalizing, and empathizing statements. Taylor's responses yielded higher rates of strategy suggestions. \ssout{}{With respect to the principles for sincere praise (P1) and process-level feedback (P2), Emily's feedback contained more statements that map onto P1, and Taylor's contained more that map onto P2.}

We describe each instructor's implementation of the GRF activity separately. For each instructor, we draw on interview data to paint an overarching picture of their perceptions about \ssout{three}{two} dimensions of their implemenations: (i) goals and framing of the GRF activity, \ssout{}{and} (ii) characteristics of good or bad feedback\ssout{, and (iii) impacts of the GRF on the nature of teacher-student relationships}{}. Then, in order to characterize the instructors' responses, we discuss the results of our GRF response coding. We focus on Emily first and Taylor second.


\subsection{Emily}

During her interview, Emily described a desire to bolster students' confidence through praise\ssout{,}{\ and to} avoid criticizing students\ssout{, and foster trusting and friendly relationships with her students}{}. Coding of GRF responses (Table~\ref{tab:codes}) revealed that encouraging and normalizing statements were each present in most of her responses. Empathizing statements, strategy suggestions, and feedback on structure were each present in about half of her responses. Resource suggestions were the least common category among her responses.

\subsubsection{Vignette: Emily's implementation}

When asked about the purpose of the reflection activity, Emily said that her goal was simply ``getting students to reflect." She said that it was important to give students an opportunity to reflect because \ssout{}{reflection is ``actually pretty important, but sometimes it's hard to set aside time in your day" to reflect.}
\SSOUT{Reflecting is something that you may just not do. \ldots\ It's actually pretty important, but sometimes it's hard to set aside time in your day or in your life to step back and reflect on what you're doing. So I guess using [the GRF] in the class was a way to give students---like, you have to reflect on your life---almost forcing them to make time to reflect. Then maybe it can become a habit later.}{Emily}
Emily said she hoped students would develop a habit of reflection that would help them avoid ``repeating things that [they] don't necessarily want to repeat" and ``engaging in some behavior that's not actually productive or helpful." When asked what she hoped her students gained from the activity, \ssout{Emily said,}{}
\SSOUT{I guess the ability to reflect on the good things they do every week---and not necessarily the negatives---because I know [the negatives are] pretty hard to not focus on. \ldots\ I hope that during those reflections, and them thinking about the good stuff that they did, helps with their confidence a little bit.}{Emily}
\ssout{Hence another of Emily's goals}{Emily said that another of her} goals for the reflection activity was to boost students' confidence by giving them opportunities to reflect on ``the good things they do every week." \ssout{Emily articulated a belief that building confidence is especially important for students from underrepresented groups studying physics:}{}
\SSOUT{It sucks, but you have to be fairly confident about your ability to do science if you want to succeed in science, especially as someone from an underrepresented group. I feel like those reflections are a way to potentially build confidence because you're reflecting on things you did well and where you need to improve upon, instead of just focusing on what you did poorly. In physics, it can be really easy to just think about what you did poorly. It's extra important to build the confidence of [students from underrepresented groups] because it's a lot easier to get your confidence crushed down, I think.}{Emily}

Emily's first goal---getting students to reflect---informed how she framed the activity to students: Emily said she told students the GRF was ``an opportunity for y'all to reflect on what you're doing." This message was communicated to the students verbally once at beginning of the semester and twice more over the duration of the course. Her second goal---boosting students' confidence---informed the type of feedback she provided.

When asked to comment on connections between criticism and support, Emily highlighted an important tension in her understanding of these two concepts. For her, criticism was related to pointing out areas for improvement, but she saw it in opposition to supportive feedback, which she defined as ``always positive," ``constructive," and requiring praise. In particular, Emily saw praise as connected to improving students' confidence:
\QUOTE{Praise, to me, is like a confidence booster type-thing. Giving praise can give someone confidence to keep trying, or keep working hard. That's mostly it. Praise is meant to encourage people to keep up their good work. \ldots\ It makes it feel like you're doing something right, you're doing something okay, and you have the ability to keep doing it.}{Emily}

On the other hand, Emily described bad feedback as ``generic," ``not sincere or not personal," and/or ``not necessarily positive or encouraging." She suggested that a lack of sincerity could potentially limit the positive impacts of praise.
\QUOTE{There's probably a way that praise can also be not supportive. If it doesn't feel like it's genuine, that could be potentially not supportive praise. Usually, [praise] would be supportive, but I think it could potentially be not supportive.}{Emily}

Emily said she preferred encouragement to criticism because she perceived critical feedback as hurtful:
\QUOTE{I try to be encouraging even if they're doing something wrong. \ldots\ [I am] never super critical. I try to be really not critical because I'm a really sensitive person, so I know that it hurts to receive critical feedback, so I avoid it all costs.}{Emily}
She described the boundary between supportive and unsupportive critical feedback as ``a fairly thin line" that she tries to ``stay above, towards the supportive end." Nevertheless, she acknowledged that criticism and support ``interact in very complex ways,'' and that students' learning can be supported by critical feedback, or hindered by its omission.

Despite seeing value in praise, Emily recognized that providing only praise may not always be the best strategy: ``If you're not critical and you constantly say, `Close! Good job!,' then they may not try to improve or learn as much." She  drew connections between lack of critical feedback, insincere praise, and the nature of student-teacher relationships:
\QUOTE{I feel like not being critical and being supportive can really enhance student-teacher relationships, but I also feel like it could potentially hurt the student-teacher relationship if you aren't being critical to the point that students start to not learn. If I'm being supportive and not critical, but then you start doing poorly in my class, you're not going to like me as much because it's like, `Why are you letting me do poorly in this class while you tell me I'm going a good job?' It's a very delicate area.}{Emily}
Emily's description of the balance between praise, criticism, and support as a ``delicate area" highlights the tension she perceived in trying to help students build confidence as learners \ssout{, support}{while supporting} them to grow in their areas of weakness\ssout{, and foster meaningful student-teacher relationships}.
 
\ssout{Consistent with her desire to provide supportive, constructive, and positive feedback, Emily also described a desire to establish supportive, trusting, and friendly relationships with her students. For example, when asked to describe her ideal student-teacher relationship, she said,}{}
\SSOUT{I feel like it would be that [students] can trust me, to come to me if they have any issues or complaints or need help with some emotional issue or something like that. That's really important to me.}{Emily}
\ssout{Moreover, one of Emily's most memorable moments from teaching Foundations involved a friendly exchange between Emily and one of her students:}{}
\SSOUT{One of the students came up to me [outside of class] \ldots\ and said that she really, really liked the feedback and appreciated what I wrote. She wrote a lot, and I also wrote a lot. I was like, `Yeah, I feel like I'm talking to one of my friends when I'm writing your feedback.' She'd share a lot, and then I'd share, too, and I feel like I got to know her really well because of those. That was really memorable for me.}{Emily}
\ssout{Emily suggested that the GRF activity played ``a big role" in helping establish similarly friendly student-teacher relationships with some other students as well. However, she also described an unanticipated barrier to the type of sharing and relationship-building she was trying to achieve:}{}
\SSOUT{I didn't sign my feedback. I kind of assumed [the students] would know [I was writing the feedback], but they didn't necessarily. This is something I saw in someone's reflection. They were upset that they didn't know who was reading them, because they didn't know how much they could share. \ldots\ I guess that should have probably been made more explicit. `Just me and [Taylor] are reading your reflections, and that's it.'}{Emily}
\ssout{This example highlights how small details---like instructors signing their feedback---can have significant impacts on students' engagement with the GRF.}{}

\subsubsection{Coding results: Emily's responses}

As can be seen in Table~\ref{tab:codes}, almost all of Emily's responses included encouraging statements. Such statements were short, and the vast majority were exclamatory. Some were motivational in nature (e.g., ``Woo! Yeah! That's the spirit!"), while others served to validate students' actions (e.g., ``I'm so glad you're working on managing your time!").

The majority of Emily's responses included normalizing statements. Emily sometimes normalized a student's experience by telling a personal anecdote from her own life that mirrored the student's experience. For example, in response to a reflection in which a student described being overwhelmed by an unexpectedly heavy workload for an Organic Chemistry course, Emily said,
\QUOTE{I had a similar experience last year. The first few weeks of my quantum mechanics class was going over such easssyyy stuff (in my opinion) \ldots\ Then all of a sudden I got slapped in the face with new material I hadn't seen before and the pace of the class started speeding up; I had a horrible time trying to catch up. I eventually did though.}{Emily}
Similarly, in response to a student who struggled with getting enough sleep and who slept through a Computer Science course, Emily said,
\QUOTE{I can totally relate! A few weeks ago I stayed up until 3am to finish an assignment and then slept through the class the assignment was for! I definitely needed the sleep though. Since then I've been trying to plan ahead a little better to avoid such late nights.}{Emily}
In each of these case, Emily drew upon a recent example from her time in graduate school in order to normalize her students' experiences.

For Emily, normalizing and empathizing often happened at the same time. Moreover, just as in the case of normalizing statements, some of Emily's empathizing statements also incorporated personal anecdotes. For example, one student described a particularly difficult lab activity that required knowing advanced Calculus content, with which the student was unfamiliar. The student concluded their reflection by saying, ``There may have been panic and tears involved." Emily responded as follows:
\QUOTE{Ahhh. I'm sorry for the panic and tears! I've definitely experienced the tears! That sounds like it would be very stressful. It sucks they didn't really prepare you for that lab.}{Emily}
Emily simultaneously normalized crying while also acknowledging that the situation described by her student was unfortunate and may have been stressful. In total, about half of Emily's responses included an empathizing statement.

Emily suggested strategies in about half of her responses to students. Some of Emily's suggestions were indirect, in the form of a personal anecdote:
\QUOTE{I use a planner to write out my tasks for the day and use [an online] calendar to keep my schedule. I have the calendar synced with my phone, so I can look at it whenever I need to and it even has reminders if I want them!}{Emily}
In other cases, Emily was more direct. For example, one student wrote that, when they can't solve a problem, they often put themselves down. The student said they wanted to improve by no longer allowing setbacks to define how they feel about themselves. To this end, Emily responded with the following suggestion:
\QUOTE{Try thinking of yourself as one of your friends. How would you react if someone was saying the putdowns you use on yourself to one of your friends? You should have that same reaction when you use those on yourself. `My friend isn't dumb. My friend is super awesome and can work hard to overcome this.'}{Emily}

Emily provided feedback about the structure of students' reflections in about half of her responses. Her feedback was often in the form of a question that, if answered, would result in a reflection that addressed the GRF responses in a more comprehensive way (e.g., ``Was there a specific experience that made you want to be more organized?," and ``What, specifically, could you do to improve your organizational skills?"). In some cases, Emily's comments about structure were formulated as requests: ``Please answer all of the questions in the reflection. Completing each question is helpful not only for the instructors but also for you."

Compared to other response types, resource suggestions were the least common for Emily. When suggesting resources, Emily typically recommended that students use on-campus tutoring services and study rooms as well as online resources. She also framed the Foundations teachers (including herself) as a resource. For example, when a student described struggling with a hard physics problem, Emily offered to help: ``If you aren't completely tired of thinking about it, I'd be happy to talk to you about the roller coaster problem and what specifically is tripping you up." Consistent with her framing of the GRF activity, Emily's primary focus was on supporting students through positive, personal connections.


\subsection{Taylor}

During his interview, Taylor said he wanted students to learn how to reflect on themselves from an objective perspective, and that he aspired to provide concrete suggestions to students about how to improve. Coding of GRF responses (Table~\ref{tab:codes}) revealed that encouraging statements and strategy statements were each present in most of his responses. About half of his responses included feedback on structure, about a third included normalizing statements, and about a quarter included resource suggestions. Empathizing statements were the least common category among his responses.

\subsubsection{Vignette: Taylor's implementation}

Taylor described three major goals for the GRF. One of Taylor's goals was for students to reflect on and improve their learning, organizational skills, and mindset: ``I hope that \ldots\ they get this growth mindset, and that they take away 1 to 2 learning strategies that we gave them." Throughout the interview, Taylor frequently related development of reflection skills to development of a growth mindset.  Another of his goals was for students to learn \emph{how} to reflect---in particular, how to do so \ssout{objectively:}{``from an objective perspective."}
\SSOUT{It's probably more important that they learn how to reflect than how good an individual reflection is. If they're able to constantly reevaluate or take a step back from themselves and look at [themselves] from an objective perspective \ldots\ they remove the frustration and emotional component out of their success. They're like, `Okay, I can do that, but only if I keep a cool head or a clear mind.' That's one of the goals of the reflections.}{Taylor}
The third goal described by Taylor was related to the creation of an avenue for communication between students and instructors\ssout{:}{, through which students ``can voice problems" and instructors can know ``what's going on with the students."}
\SSOUT{Another [goal] is that the instructors know much better what's going on with the students. Also the students have some confidential space where they can voice problems that they see.}{Taylor}

Taylor's focus on students' ability to take an objective perspective on their own learning and his goal of creating confidential communication pathways between students and teachers informed how he framed the activity to his students\ssout{:}{. He said that he told students that it is important be ``able to take an objective perspective on yourself" and to communicate with instructors ``in a very private, confidential space that is not rushed or in somebody's office."}
\SSOUT{What I basically tried to tell them is, `What this is important for is, you are able to take an objective perspective on yourself and also a perspective that makes you a better learner. Furthermore it allows you to communicate with us in a very private, confidential space that is not rushed or in somebody's office.'}{Taylor}
Taylor said he communicated this framing to his class verbally at the start of the semester, and that he reinforced this framing throughout the duration of the course in his written feedback to students and in one-on-one conversations with students.

Taylor described good feedback as ``logical," ``concrete," and ``to the point," whereas bad feedback was described as ``confusing," ``negative," and not valuing the students' effort. Taylor expressed concern about praise that focused on students' inherited traits or that was insincere. He noted that, in addition to focusing on students' effort, praise should also be tailored to the circumstances of the particular student being praised: 
\QUOTE{You should praise definitely the effort, their attitude towards working, [rather than] the things they inherited from wherever. The other thing is, if you give a lot of praise \ldots\ praise no longer becomes genuine. It just becomes some sort of mechanism. You should always try to have some personal note in there. \ldots\ Praise should be individual, and it should be appropriate.}{Taylor}
Taylor also described providing feedback to help students develop specific study habits:
\QUOTE{I also gave them strategies how to change their learning schedule. There are studies that after 45 minutes it's essentially pointless to go on, you should have a short break where your brain can regenerate. I definitely wrote that on every single reflection about learning skills. A few students actually responded the following weeks that they started doing that and saw gains in their learning.}{Taylor}

However, \ssout{he}{Taylor} noted that it can be difficult to provide good feedback to students who write short reflections that lack specificity:
\QUOTE{If you write something very general, then you can use very little words to describe a lot of situations. But the devil is in the details, and it's difficult to give the student appropriate feedback. \ldots\ If you hit a certain low word count, you just cannot say a lot of things. Normally, low word count goes along with very general statements, and that's hard to give feedback on.}{Taylor}
\ssout{According to Taylor, vague reflections were not conducive to concrete (i.e., good) feedback. When discussing student-teacher relationships, Taylor emphasized the difference between friend and teacher:}{}
\SSOUT{[The students] were able to see me as an ally. Of course not friend, because I'm still their instructor or teacher. \ldots\ I think this is just great, because you can have a personal relationship but still work with them as a teacher.}{Taylor}
\ssout{Consistent with his interpretation of the GRF as a communication avenue between students and instructors, Taylor said that the activity gave him access to a type of working relationship with his students that he normally doesn't have access to:}{}
\SSOUT{The reflections allowed me to sometimes personally address problems with the students. \ldots\ This was very helpful to establish a good working relationship with the students. This is something I normally don't have access to, but now for some students I had access to.}{Taylor}
\ssout{However, Taylor was not able to use the GRF to establish this type of rapport with every student.}{} When asked to describe something he found surprising about the course, Taylor recounted an experience with a student who wrote short reflections throughout the semester: ``There was one student where I could not achieve that the student wrote long reflections." Taylor said he tried to encourage this student to write longer reflections both in his written feedback on the GRF as well as verbally during class. These efforts did not work, which surprised Taylor:
\QUOTE{That was a little bit surprising because normally students always have a lot of things to tell, and these are the things that normally nobody talks about with them. So it was somewhat surprising.}{Taylor}
For Taylor, \ssout{not only did short reflections make it difficult to write good feedback, they were also perceived as a barrier to connecting with students}{short or vague reflections made it difficult to write good (i.e., concrete) feedback}. Indeed, upon reading a draft of this manuscript, Taylor asked us to emphasize that short reflections were ``one of the worst obstacles" to productive use of the GRF.

\subsubsection{Coding results: Taylor's responses}

As can be seen in Table~\ref{tab:codes}, Taylor included encouraging statements in a majority of his responses. Many of these statements were one-word exclamations (e.g., ``Great!," and ``Nice!"). Less often, Taylor also validated students' actions: ``It is very good that you acknowledge the importance of physical and psychological well-being by taking a break from work."

A majority of Taylor's responses included strategy suggestions. Taylor's suggestions were often straightforward and direct. For example, one student described having difficulty during a group activity where some group members could not reach agreement about whether light is displaced or bent as is passes through a medium. The student said they wished they could communicate with their group. In response, Taylor recommended using an alternative mode of communication:
\QUOTE{Have you thought about different ways to communicate? Since it was an optical phenomenon, maybe drawings could help you to communicate your thoughts, to also solidify them for yourself.}{Taylor}
Sometimes, Taylor used anecdotes to make indirect suggestions. In response to a student who described the balance between attending class, studying, and having time to unwind as stressful and strenuous, Taylor said,
\QUOTE{Think also about your physical and psychological well-being by giving yourself (short) breaks from studying. I myself go for a short walk every day just to clear my head and get some fresh air.}{Taylor}
Here, Taylor's description of his own strategy---going for short walks---served as an indirect suggestion to the student.

About half of Taylor's responses included feedback on the structure of the reflection itself. Sometimes this feedback was formulated as a question (e.g., ``What is a concrete strategy you want to use?"), but more often it was a direct request (e.g., ``I still would like to encourage you to write more, so I can give you better feedback."). Taylor's structure feedback often focused on encouraging students to write longer and more specific reflections. Another common theme was reminding students that their reflections could focus on experiences outside of the context of the Foundations course. In the following example, Taylor provided all of these types of structure feedback:
\QUOTE{Maybe you can expand your responses a little bit and become more concrete in the answers, e.g., what exact strategy you want to employ, or what particular obstacle you faced during the class section. Also, this reflection is not just limited to your experience in class, but about all your classes!}{Taylor}
Sometimes, Taylor was successful in getting a student to write more; in these cases, Taylor provided feedback acknowledging the improvement (e.g., ``Your reflection has significantly improved.").

Fewer than half of Taylor's responses included normalizing and/or empathizing statements. When normalizing students' experiences, Taylor typically framed those experiences as common or typical (e.g., ``Many students feel that way," and ``Everybody needs some outside help."). And when empathizing with students, Taylor often focused on acknowledging their feelings (e.g., ``It seems that you were overwhelmed by the plethora of tasks," and ``This is a pretty terrible experience you described above.").

Resource suggestions were present in about a quarter of Taylor's responses. Like Emily, Taylor also recommended that students make use of on-campus tutoring services and study rooms, online resources, and the Foundations teachers (including himself). In addition, Taylor often recommended specific books where students could find additional practice problems:
\QUOTE{There is a plethora of really good books with various difficulty levels of problems \ldots\ Please ask me for more if you think that this would be useful for you!}{Taylor}
(Note that we have omitted from the quote the names of the particular books Taylor recommended.) In this example, Taylor not only recommended specific books as a resource, but also offered himself as a resource that the student could turn to for additional recommendations.


\section{Summary and Discussion}\label{sec:discussion}

Our overarching goal for the GRF is to provide avenues through which students can receive personal attention from instructors about a variety of challenges they may experience while learning physics---interactions that are discouraged by weed-out culture.\cite{Seymour1997} Our vision for its use \ssout{}{is} informed by Brown's metaphor for feedback:\cite{Brown2012} instructors and students ``sitting on the same side of the table" while working together to improve students' learning experiences. As a step toward understanding whether and how the GRF is able to support these types of interactions, we performed an exploratory qualitative study of two graduate student instructors' implementations of the GRF in an introductory lab course whose learning goals and broader programmatic context emphasized developing students' reflection skills\ssout{and fostering supportive student-teacher relationships}{}. This learning environment not only resonated with our vision for how the GRF might ideally be used by physics instructors, it also aligned with characteristics of environments in which students are well situated to receive and use feedback.\cite{Brookhart2008} 

Our analysis drew on two sources of data: (i) post-semester interviews with Emily and Taylor and (ii) their written responses to 134 student reflections. We found that Emily and Taylor \ssout{both perceived that the GRF played an important role in the formation of meaningful connections with their students, and they}{} each used all six of the following response types: encouraging statements, normalizing statements, empathizing statements, strategy suggestions, resource suggestions, and feedback to the student on the structure of their reflection. In particular, strategy suggestions were present in about half of Emily's responses and in most of Taylor's; this type of process-level feedback is especially effective for improving students' learning.\cite{Hattie2007}

\ssout{Within CU-Prime, building community among graduate and undergraduate students is an explicit goal of many programs---including the Foundations course. As such, there is an inherent tension between maintaining teacher-student boundaries and cultivating friendships in Foundations. Taylor and Emily each navigated this tension differently. Both Emily and Taylor said that the reflection activity enabled them to get to know their students well and facilitated the development of personal relationships with their students that extended beyond the course context. Emily said that GRF facilitated two-directional sharing, which in turn fostered a friendship-style relationship with one of her students. In her responses, she often shared personal anecdotes as a way to empathize with students and/or normalize their experiences. Taylor, on the other hand, described a desire to establish a good working relationship with his students. He formatted his responses as if he were writing letters to his students, and he focused on suggesting concrete strategies that students could use to improve their learning habits. Thus, through the GRF, each instructor was able to establish a distinct balance between teacher and friend with which they were satisfied.}{}

According to Seymour and Hewitt,\cite{Seymour1997} weed-out culture encourages students to ``cast aside dependence on personally-significant adults and take responsibility for their own learning\ssout{" (p.~132).}{."} We argue that, counter to this culture, the GRF supported Emily and Taylor to simultaneously position themselves as people on whom students could depend \emph{and} encourage students to take responsibility for their own learning. For example, by offering to help students solve a roller-coaster problem or suggesting books with useful practice problems, Emily and Taylor framed themselves as resources for students' coursework outside the context of Foundations. Meanwhile, the instructors aimed to promote self-regulation and confidence by suggesting strategies through which students could grow as learners, as well as providing feedback on how to engage in the act of reflection itself.

Instructors' feedback on the structure of students' reflections focused on the specificity, completeness, and length of reflections. In some cases, the instructors were successful in getting students to improve the quality of their reflections along these metrics. In other cases, however, they were not. During their interviews, both instructors said that short reflections made it difficult to write good feedback and/or connect with the student. However, we note that not all students may need or want to engage in the type of student-teacher interaction facilitated by the GRF. While we did not see evidence that the instructors were imposing the GRF on any of the Foundations students, we caution that doing so would be counter to the spirit of students and teachers (voluntarily) ``sitting on the same side of the table" to tackle hard problems together. It can be difficult to know which students will respond to structure feedback and which will ignore it. The tension between improving engagement with the GRF among some students while respecting that others may not want to engage at all is an important area for future investigation.

On a related note, some students may choose to engage with the GRF in only cursory ways (or not at all) because it erroneously assumes that they experience something upon which they would like to improve on a weekly basis. The prompts may also unintentionally situate challenging experiences as failures. For example, although we did not analyze student reflections in this work, we found one excerpt from a Foundations student particularly insightful. In week 7, the student modified the GRF prompts related to resilience before reflecting on a health issue they were experiencing. At the bottom of their reflection, the student explained why they modified the prompts:
\QUOTE{I removed all mention of the word `failure.'  \ldots\ I NEVER would have written this story on the original form. Why? Because it is NOT a story of failure, or even `something that I would like to improve upon' (original prompt \#1). Potentially having a [health issue] is NOT a failure on my part \ldots\ It's just a challenging situation.}{C06-7}
This suggests that engagement in the GRF could be improved by making changes to both the framing and language of the activity.

We hope this paper can be a resource to physics instructors who are using the GRF or similar activities in their classrooms, especially those who are part of CU-Prime or other organizations within The Access Network and may therefore have similar learning goals and environments to Foundations. Together with previous work on structure of student reflections, this work lays the foundation for future research about the ways that feedback and reflections can be mutually informative for one another, and how the GRF impacts  student-teacher interactions (short-term) and student retention (long-term). \ssout{}{In particular, the impact of the GRF on persistence of physics students from marginalized populations is an important topic for future investigation. Such studies will become possible in the coming years, as the first cohorts of students who completed Foundations begin graduating from college. More generally, we hope that these findings inform the implementation of the GRF in courses with small student-to-teacher ratios, or in large courses with small sections led by teaching or learning assistants. The GRF could also be implemented in informal educational contexts, including academic advising or mentorship programs.}


\acknowledgments This material is based upon work supported by the NSF under Grant Nos. DUE-1323101 and PHY-1125844. Both authors contributed equally to this work.

\bibliography{database_TAG-GRF}

\begin{thebibliography}{26}%
\makeatletter
\providecommand \@ifxundefined [1]{%
 \@ifx{#1\undefined}
}%
\providecommand \@ifnum [1]{%
 \ifnum #1\expandafter \@firstoftwo
 \else \expandafter \@secondoftwo
 \fi
}%
\providecommand \@ifx [1]{%
 \ifx #1\expandafter \@firstoftwo
 \else \expandafter \@secondoftwo
 \fi
}%
\providecommand \natexlab [1]{#1}%
\providecommand \enquote  [1]{``#1''}%
\providecommand \bibnamefont  [1]{#1}%
\providecommand \bibfnamefont [1]{#1}%
\providecommand \citenamefont [1]{#1}%
\providecommand \href@noop [0]{\@secondoftwo}%
\providecommand \href [0]{\begingroup \@sanitize@url \@href}%
\providecommand \@href[1]{\@@startlink{#1}\@@href}%
\providecommand \@@href[1]{\endgroup#1\@@endlink}%
\providecommand \@sanitize@url [0]{\catcode `\\12\catcode `\$12\catcode
  `\&12\catcode `\#12\catcode `\^12\catcode `\_12\catcode `\%12\relax}%
\providecommand \@@startlink[1]{}%
\providecommand \@@endlink[0]{}%
\providecommand \url  [0]{\begingroup\@sanitize@url \@url }%
\providecommand \@url [1]{\endgroup\@href {#1}{\urlprefix }}%
\providecommand \urlprefix  [0]{URL }%
\providecommand \Eprint [0]{\href }%
\providecommand \doibase [0]{http://dx.doi.org/}%
\providecommand \selectlanguage [0]{\@gobble}%
\providecommand \bibinfo  [0]{\@secondoftwo}%
\providecommand \bibfield  [0]{\@secondoftwo}%
\providecommand \translation [1]{[#1]}%
\providecommand \BibitemOpen [0]{}%
\providecommand \bibitemStop [0]{}%
\providecommand \bibitemNoStop [0]{.\EOS\space}%
\providecommand \EOS [0]{\spacefactor3000\relax}%
\providecommand \BibitemShut  [1]{\csname bibitem#1\endcsname}%
\let\auto@bib@innerbib\@empty
\bibitem [{\citenamefont {Ward}\ and\ \citenamefont {Duda}(2014)}]{Ward2014}%
  \BibitemOpen
  \bibfield  {author} {\bibinfo {author} {\bibfnamefont {K.}~\bibnamefont
  {Ward}}\ and\ \bibinfo {author} {\bibfnamefont {G.}~\bibnamefont {Duda}},\
  }\href
  {http://www.compadre.org/PER/perc/2014/files/Ward_Duda_PERC2014_reflection_ver3.pdf}
  {\emph {\bibinfo {title} {The role of student reflection in project-based
  learning physics courses}}}\ (\bibinfo  {publisher} {Proceedings of the 2014
  Physics Education Research Conference},\ \bibinfo {address} {Minneapolis,
  MN},\ \bibinfo {year} {2014})\BibitemShut {NoStop}%
\bibitem [{\citenamefont {Mason}\ and\ \citenamefont
  {Singh}(2010)}]{Mason2010}%
  \BibitemOpen
  \bibfield  {author} {\bibinfo {author} {\bibfnamefont {Andrew}\ \bibnamefont
  {Mason}}\ and\ \bibinfo {author} {\bibfnamefont {Chandralekha}\ \bibnamefont
  {Singh}},\ }\bibfield  {title} {\enquote {\bibinfo {title} {Using reflection
  with peers to help students learn effective problem solving strategies},}\
  }\href {\doibase http://dx.doi.org/10.1063/1.3515243} {\bibfield  {journal}
  {\bibinfo  {journal} {AIP Conference Proceedings}\ }\textbf {\bibinfo
  {volume} {1289}},\ \bibinfo {pages} {41--44} (\bibinfo {year}
  {2010})}\BibitemShut {NoStop}%
\bibitem [{\citenamefont {Etkina}\ \emph {et~al.}(2010)\citenamefont {Etkina},
  \citenamefont {Karelina}, \citenamefont {Ruibal-Villasenor}, \citenamefont
  {Rosengrant}, \citenamefont {Jordan},\ and\ \citenamefont
  {Hmelo-Silver}}]{Etkina2010}%
  \BibitemOpen
  \bibfield  {author} {\bibinfo {author} {\bibfnamefont {Eugenia}\ \bibnamefont
  {Etkina}}, \bibinfo {author} {\bibfnamefont {Anna}\ \bibnamefont {Karelina}},
  \bibinfo {author} {\bibfnamefont {Maria}\ \bibnamefont {Ruibal-Villasenor}},
  \bibinfo {author} {\bibfnamefont {David}\ \bibnamefont {Rosengrant}},
  \bibinfo {author} {\bibfnamefont {Rebecca}\ \bibnamefont {Jordan}}, \ and\
  \bibinfo {author} {\bibfnamefont {Cindy~E.}\ \bibnamefont {Hmelo-Silver}},\
  }\bibfield  {title} {\enquote {\bibinfo {title} {Design and reflection help
  students develop scientific abilities: Learning in introductory physics
  laboratories},}\ }\href {\doibase 10.1080/10508400903452876} {\bibfield
  {journal} {\bibinfo  {journal} {Journal of the Learning Sciences}\ }\textbf
  {\bibinfo {volume} {19}},\ \bibinfo {pages} {54--98} (\bibinfo {year}
  {2010})}\BibitemShut {NoStop}%
\bibitem [{\citenamefont {Scott}\ \emph {et~al.}(2007)\citenamefont {Scott},
  \citenamefont {Stelzer},\ and\ \citenamefont {Gladding}}]{Scott2007}%
  \BibitemOpen
  \bibfield  {author} {\bibinfo {author} {\bibfnamefont {Michael~L.}\
  \bibnamefont {Scott}}, \bibinfo {author} {\bibfnamefont {Tim}\ \bibnamefont
  {Stelzer}}, \ and\ \bibinfo {author} {\bibfnamefont {Gary}\ \bibnamefont
  {Gladding}},\ }\bibfield  {title} {\enquote {\bibinfo {title} {Explicit
  reflection in an introductory physics course},}\ }\href {\doibase
  http://dx.doi.org/10.1063/1.2820929} {\bibfield  {journal} {\bibinfo
  {journal} {AIP Conference Proceedings}\ }\textbf {\bibinfo {volume} {951}},\
  \bibinfo {pages} {188--191} (\bibinfo {year} {2007})}\BibitemShut {NoStop}%
\bibitem [{\citenamefont {May}\ and\ \citenamefont {Etkina}(2002)}]{May2002}%
  \BibitemOpen
  \bibfield  {author} {\bibinfo {author} {\bibfnamefont {David~B.}\
  \bibnamefont {May}}\ and\ \bibinfo {author} {\bibfnamefont {Eugenia}\
  \bibnamefont {Etkina}},\ }\bibfield  {title} {\enquote {\bibinfo {title}
  {College physics students' epistemological self-reflection and its
  relationship to conceptual learning},}\ }\href {\doibase
  http://dx.doi.org/10.1119/1.1503377} {\bibfield  {journal} {\bibinfo
  {journal} {American Journal of Physics}\ }\textbf {\bibinfo {volume} {70}},\
  \bibinfo {pages} {1249--1258} (\bibinfo {year} {2002})}\BibitemShut {NoStop}%
\bibitem [{\citenamefont {{National Academies of Sciences, Engineering, and
  Medicine}}(2015)}]{NASEM2015}%
  \BibitemOpen
  \bibfield  {author} {\bibinfo {author} {\bibnamefont {{National Academies of
  Sciences, Engineering, and Medicine}}},\ }\href
  {http://www.nap.edu/catalog/21851/integrating-discovery-based-research-into-the-undergraduate-curriculum-report-of}
  {\emph {\bibinfo {title} {Integrating Discovery-Based Research into the
  Undergraduate Curriculum: Report of a Convocation}}}\ (\bibinfo  {publisher}
  {National Academies Press},\ \bibinfo {year} {2015})\ Chap.\ \bibinfo
  {chapter} {{3, Promising Practices and Ongoing Challenges}}\BibitemShut
  {NoStop}%
\bibitem [{\citenamefont {Zimmerman}(2002)}]{Zimmerman2002}%
  \BibitemOpen
  \bibfield  {author} {\bibinfo {author} {\bibfnamefont {Barry~J.}\
  \bibnamefont {Zimmerman}},\ }\bibfield  {title} {\enquote {\bibinfo {title}
  {Becoming a self-regulated learner: An overview},}\ }\href {\doibase
  10.1207/s15430421tip4102_2} {\bibfield  {journal} {\bibinfo  {journal}
  {Theory Into Practice}\ }\textbf {\bibinfo {volume} {41}},\ \bibinfo {pages}
  {64--70} (\bibinfo {year} {2002})}\BibitemShut {NoStop}%
\bibitem [{\citenamefont {Reinholz}\ and\ \citenamefont
  {Dounas-Frazer}(2016)}]{Reinholz2016}%
  \BibitemOpen
  \bibfield  {author} {\bibinfo {author} {\bibfnamefont {Daniel~L.}\
  \bibnamefont {Reinholz}}\ and\ \bibinfo {author} {\bibfnamefont {Dimitri~R.}\
  \bibnamefont {Dounas-Frazer}},\ }\bibfield  {title} {\enquote {\bibinfo
  {title} {Using peer feedback to promote reflection on open-ended problems},}\
  }\href {\doibase http://dx.doi.org/10.1119/1.4961181} {\bibfield  {journal}
  {\bibinfo  {journal} {The Physics Teacher}\ }\textbf {\bibinfo {volume}
  {54}},\ \bibinfo {pages} {364--368} (\bibinfo {year} {2016})}\BibitemShut
  {NoStop}%
\bibitem [{\citenamefont {Gandhi}\ \emph {et~al.}(2016)\citenamefont {Gandhi},
  \citenamefont {Livezey}, \citenamefont {Zaniewski}, \citenamefont
  {Reinholz},\ and\ \citenamefont {Dounas-Frazer}}]{Gandhi2016}%
  \BibitemOpen
  \bibfield  {author} {\bibinfo {author} {\bibfnamefont {Punit~R.}\
  \bibnamefont {Gandhi}}, \bibinfo {author} {\bibfnamefont {Jesse~A.}\
  \bibnamefont {Livezey}}, \bibinfo {author} {\bibfnamefont {Anna~M.}\
  \bibnamefont {Zaniewski}}, \bibinfo {author} {\bibfnamefont {Daniel~L.}\
  \bibnamefont {Reinholz}}, \ and\ \bibinfo {author} {\bibfnamefont
  {Dimitri~R.}\ \bibnamefont {Dounas-Frazer}},\ }\bibfield  {title} {\enquote
  {\bibinfo {title} {Attending to experimental physics practices and lifelong
  learning skills in an introductory laboratory course},}\ }\href@noop {}
  {\bibfield  {journal} {\bibinfo  {journal} {American Journal of Physics}\
  }\textbf {\bibinfo {volume} {84}} (\bibinfo {year} {2016})}\BibitemShut
  {NoStop}%
\bibitem [{\citenamefont {Dounas-Frazer}\ and\ \citenamefont
  {Reinholz}(2015)}]{Dounas-Frazer2015}%
  \BibitemOpen
  \bibfield  {author} {\bibinfo {author} {\bibfnamefont {Dimitri~R.}\
  \bibnamefont {Dounas-Frazer}}\ and\ \bibinfo {author} {\bibfnamefont
  {Daniel~L.}\ \bibnamefont {Reinholz}},\ }\bibfield  {title} {\enquote
  {\bibinfo {title} {Attending to lifelong learning skills through guided
  reflection in a physics class},}\ }\href {\doibase
  http://dx.doi.org/10.1119/1.4930083} {\bibfield  {journal} {\bibinfo
  {journal} {American Journal of Physics}\ }\textbf {\bibinfo {volume} {83}},\
  \bibinfo {pages} {881--891} (\bibinfo {year} {2015})}\BibitemShut {NoStop}%
\bibitem [{\citenamefont {Seymour}\ and\ \citenamefont
  {Hewitt}(1997)}]{Seymour1997}%
  \BibitemOpen
  \bibfield  {author} {\bibinfo {author} {\bibfnamefont {Elaine}\ \bibnamefont
  {Seymour}}\ and\ \bibinfo {author} {\bibfnamefont {Nancy~M.}\ \bibnamefont
  {Hewitt}},\ }\href@noop {} {\emph {\bibinfo {title} {{Talking about leaving:
  Why undergraduates leave the sciences}}}}\ (\bibinfo  {publisher} {Westview
  Press},\ \bibinfo {address} {Boulder, CO},\ \bibinfo {year}
  {1997})\BibitemShut {NoStop}%
\bibitem [{\citenamefont {Brown}(2012)}]{Brown2012}%
  \BibitemOpen
  \bibfield  {author} {\bibinfo {author} {\bibfnamefont {Bren\'e}\ \bibnamefont
  {Brown}},\ }\href@noop {} {\emph {\bibinfo {title} {{Daring greatly: How the
  courage to be vulnerable transforms the way we live, love, parent, and
  lead}}}}\ (\bibinfo  {publisher} {Penguin},\ \bibinfo {address} {New York,
  NY},\ \bibinfo {year} {2012})\BibitemShut {NoStop}%
\bibitem [{\citenamefont {Shute}(2008)}]{shute2008}%
  \BibitemOpen
  \bibfield  {author} {\bibinfo {author} {\bibfnamefont {V.J.}\ \bibnamefont
  {Shute}},\ }\bibfield  {title} {\enquote {\bibinfo {title} {Focus on
  formative feedback},}\ }\href {\doibase 10.3102/0034654307313795} {\bibfield
  {journal} {\bibinfo  {journal} {Review of Educational Research}\ }\textbf
  {\bibinfo {volume} {78}},\ \bibinfo {pages} {153--189} (\bibinfo {year}
  {2008})}\BibitemShut {NoStop}%
\bibitem [{\citenamefont {Hattie}\ and\ \citenamefont
  {Timperley}(2007)}]{Hattie2007}%
  \BibitemOpen
  \bibfield  {author} {\bibinfo {author} {\bibfnamefont {John}\ \bibnamefont
  {Hattie}}\ and\ \bibinfo {author} {\bibfnamefont {Helen}\ \bibnamefont
  {Timperley}},\ }\bibfield  {title} {\enquote {\bibinfo {title} {The power of
  feedback},}\ }\href {\doibase 10.3102/003465430298487} {\bibfield  {journal}
  {\bibinfo  {journal} {Review of Educational Research}\ }\textbf {\bibinfo
  {volume} {77}},\ \bibinfo {pages} {81--112} (\bibinfo {year}
  {2007})}\BibitemShut {NoStop}%
\bibitem [{\citenamefont {Rattan}\ \emph {et~al.}(2012)\citenamefont {Rattan},
  \citenamefont {Good},\ and\ \citenamefont {Dweck}}]{Rattan2012}%
  \BibitemOpen
  \bibfield  {author} {\bibinfo {author} {\bibfnamefont {Aneeta}\ \bibnamefont
  {Rattan}}, \bibinfo {author} {\bibfnamefont {Catherine}\ \bibnamefont
  {Good}}, \ and\ \bibinfo {author} {\bibfnamefont {Carol~S.}\ \bibnamefont
  {Dweck}},\ }\bibfield  {title} {\enquote {\bibinfo {title} {{``It's ok---Not
  everyone can be good at math": Instructors with an entity theory comfort (and
  demotivate) students}},}\ }\href {\doibase
  http://dx.doi.org/10.1016/j.jesp.2011.12.012} {\bibfield  {journal} {\bibinfo
   {journal} {Journal of Experimental Social Psychology}\ }\textbf {\bibinfo
  {volume} {48}},\ \bibinfo {pages} {731 -- 737} (\bibinfo {year}
  {2012})}\BibitemShut {NoStop}%
\bibitem [{\citenamefont {Dweck}(2006)}]{Dweck2006}%
  \BibitemOpen
  \bibfield  {author} {\bibinfo {author} {\bibfnamefont {Carol~S.}\
  \bibnamefont {Dweck}},\ }\href@noop {} {\emph {\bibinfo {title} {{Mindset:
  The New Psychology of Success}}}}\ (\bibinfo  {publisher} {Random House},\
  \bibinfo {year} {2006})\BibitemShut {NoStop}%
\bibitem [{\citenamefont {Bandura}(1997)}]{Bandura1997}%
  \BibitemOpen
  \bibfield  {author} {\bibinfo {author} {\bibfnamefont {Albert}\ \bibnamefont
  {Bandura}},\ }\bibfield  {title} {\enquote {\bibinfo {title} {Sources of
  self-efficacy},}\ }in\ \href@noop {} {\emph {\bibinfo {booktitle}
  {{Self-efficacy: The exercise of control}}}}\ (\bibinfo  {publisher} {W. H.
  Freeman},\ \bibinfo {address} {New York, NY},\ \bibinfo {year} {1997})\ pp.\
  \bibinfo {pages} {79--115}\BibitemShut {NoStop}%
\bibitem [{\citenamefont {Brookhart}(2008)}]{Brookhart2008}%
  \BibitemOpen
  \bibfield  {author} {\bibinfo {author} {\bibfnamefont {Susan~M.}\
  \bibnamefont {Brookhart}},\ }\bibfield  {title} {\enquote {\bibinfo {title}
  {{Feedback: An overview}},}\ }in\ \href@noop {} {\emph {\bibinfo {booktitle}
  {How to give effective feedback to your students}}}\ (\bibinfo  {publisher}
  {ASCD},\ \bibinfo {address} {Alexandria, VA},\ \bibinfo {year} {2008})\ pp.\
  \bibinfo {pages} {1--30}\BibitemShut {NoStop}%
\bibitem [{\citenamefont {Eisenhart}(2009)}]{Eisenhart2009}%
  \BibitemOpen
  \bibfield  {author} {\bibinfo {author} {\bibfnamefont {Margaret}\
  \bibnamefont {Eisenhart}},\ }\bibfield  {title} {\enquote {\bibinfo {title}
  {Generalization from qualitative inquiry},}\ }in\ \href@noop {} {\emph
  {\bibinfo {booktitle} {{Generalizing from Educational Research: Beyond
  Qualitative and Quantitative Polarization}}}},\ \bibinfo {editor} {edited by\
  \bibinfo {editor} {\bibfnamefont {Kadriye}\ \bibnamefont {Ercikan}}\ and\
  \bibinfo {editor} {\bibfnamefont {Wolff-Michael}\ \bibnamefont {Roth}}}\
  (\bibinfo {address} {New York, NY},\ \bibinfo {year} {2009})\BibitemShut
  {NoStop}%
\bibitem [{Note1()}]{Note1}%
  \BibitemOpen
  \bibinfo {note} {The Access Network, \protect \url
  {http://accessnetwork.org}}\BibitemShut {NoStop}%
\bibitem [{\citenamefont {Albanna}\ \emph {et~al.}(2013)\citenamefont
  {Albanna}, \citenamefont {Corbo}, \citenamefont {Dounas-Frazer},
  \citenamefont {Little},\ and\ \citenamefont {Zaniewski}}]{Albanna2013}%
  \BibitemOpen
  \bibfield  {author} {\bibinfo {author} {\bibfnamefont {Badr~F.}\ \bibnamefont
  {Albanna}}, \bibinfo {author} {\bibfnamefont {Joel~C.}\ \bibnamefont
  {Corbo}}, \bibinfo {author} {\bibfnamefont {Dimitri~R.}\ \bibnamefont
  {Dounas-Frazer}}, \bibinfo {author} {\bibfnamefont {Angela}\ \bibnamefont
  {Little}}, \ and\ \bibinfo {author} {\bibfnamefont {Anna~M.}\ \bibnamefont
  {Zaniewski}},\ }\bibfield  {title} {\enquote {\bibinfo {title} {Building
  classroom and organizational structure around positive cultural values},}\
  }\href {\doibase 10.1063/1.4789638} {\bibfield  {journal} {\bibinfo
  {journal} {AIP Conference Proceedings}\ }\textbf {\bibinfo {volume} {1513}},\
  \bibinfo {pages} {7--10} (\bibinfo {year} {2013})}\BibitemShut {NoStop}%
\bibitem [{\citenamefont {Sabella}\ \emph {et~al.}(In review)\citenamefont
  {Sabella}, \citenamefont {Mardis}, \citenamefont {Little},\ and\
  \citenamefont {Sanders}}]{Sabella_inreview}%
  \BibitemOpen
  \bibfield  {author} {\bibinfo {author} {\bibfnamefont {Mel~S.}\ \bibnamefont
  {Sabella}}, \bibinfo {author} {\bibfnamefont {Kristy}\ \bibnamefont
  {Mardis}}, \bibinfo {author} {\bibfnamefont {Angie}\ \bibnamefont {Little}},
  \ and\ \bibinfo {author} {\bibfnamefont {Nicolette}\ \bibnamefont
  {Sanders}},\ }\href@noop {} {\enquote {\bibinfo {title} {{The Chi-Sci
  Scholars Program: Developing Community and Challenging Racially Inequitable
  Measures of Success at a Minority Serving Institution on Chicago's
  Southside}},}\ } (\bibinfo {year} {In review})\BibitemShut {NoStop}%
\bibitem [{\citenamefont {Dounas-Frazer}\ \emph {et~al.}(2013)\citenamefont
  {Dounas-Frazer}, \citenamefont {Lynn}, \citenamefont {Zaniewski},\ and\
  \citenamefont {Roth}}]{Dounas-Frazer2013TPT}%
  \BibitemOpen
  \bibfield  {author} {\bibinfo {author} {\bibfnamefont {D.~R.}\ \bibnamefont
  {Dounas-Frazer}}, \bibinfo {author} {\bibfnamefont {J.}~\bibnamefont {Lynn}},
  \bibinfo {author} {\bibfnamefont {A.~M.}\ \bibnamefont {Zaniewski}}, \ and\
  \bibinfo {author} {\bibfnamefont {N.}~\bibnamefont {Roth}},\ }\bibfield
  {title} {\enquote {\bibinfo {title} {Learning about non-{Newtonian} fluids in
  a student-driven classroom},}\ }\href {\doibase 10.1119/1.4772035} {\bibfield
   {journal} {\bibinfo  {journal} {The Physics Teacher}\ }\textbf {\bibinfo
  {volume} {51}},\ \bibinfo {pages} {32--34} (\bibinfo {year}
  {2013})}\BibitemShut {NoStop}%
\bibitem [{\citenamefont {Dounas-Frazer}\ \emph {et~al.}(In press)\citenamefont
  {Dounas-Frazer}, \citenamefont {Hyater-Adams},\ and\ \citenamefont
  {Reinholz}}]{Dounas-Frazer_inpress}%
  \BibitemOpen
  \bibfield  {author} {\bibinfo {author} {\bibfnamefont {Dimitri~R.}\
  \bibnamefont {Dounas-Frazer}}, \bibinfo {author} {\bibfnamefont {Simone~A.}\
  \bibnamefont {Hyater-Adams}}, \ and\ \bibinfo {author} {\bibfnamefont
  {Daniel~L.}\ \bibnamefont {Reinholz}},\ }\bibfield  {title} {\enquote
  {\bibinfo {title} {Learning to do diversity work: {A} model for continued
  education of program organizers},}\ }\href@noop {} {\bibfield  {journal}
  {\bibinfo  {journal} {The Physics Teacher}\ } (\bibinfo {year} {In
  press})}\BibitemShut {NoStop}%
\bibitem [{\citenamefont {Zaniewski}\ and\ \citenamefont
  {Reinholz}(2016)}]{Zaniewski2016}%
  \BibitemOpen
  \bibfield  {author} {\bibinfo {author} {\bibfnamefont {Anna~M.}\ \bibnamefont
  {Zaniewski}}\ and\ \bibinfo {author} {\bibfnamefont {Daniel}\ \bibnamefont
  {Reinholz}},\ }\bibfield  {title} {\enquote {\bibinfo {title} {Increasing
  {STEM} success: a near-peer mentoring program in the physical sciences},}\
  }\href {\doibase 10.1186/s40594-016-0043-2} {\bibfield  {journal} {\bibinfo
  {journal} {International Journal of STEM Education}\ }\textbf {\bibinfo
  {volume} {3}},\ \bibinfo {pages} {14} (\bibinfo {year} {2016})}\BibitemShut
  {NoStop}%
\bibitem [{GRF()}]{GRFguidelines}%
  \BibitemOpen
  \href@noop {} {}\bibinfo {note} {GRF Guidelines to be submitted as
  supplementary materials; temporary link for review:
  \url{http://tinyurl.com/Dounas-FrazerEPAPS-2017}}\BibitemShut {NoStop}%
\end{thebibliography}%

\end{document}